\documentstyle[prl, aps, amssymb
]{revtex}
\begin{document}
\draft
\wideabs{
\title{Novel Spin-Gap Behavior in Layered $S=\frac{1}{2}$ Quantum Spin System TiOCl}
\author{T. Imai$^{a}$ and F.C. Chou$^{b}$}
\address{$^{a}$Department of Physics and Astronomy,
McMaster University, Hamilton, Ontario L8S 4M1, CANADA}
\address{$^{b}$Center for Materials Science and 
Engineering, M.I.T., Cambridge, MA 02139, U.S.A.}
\date{\today}
\maketitle
\begin{abstract}
We have investigated the spin fluctuations and their 
interplay with lattice instabilities in
a layered $S=\frac{1}{2}$ quantum 
spin system TiOCl.  Our $^{47,49}$Ti and $^{35}$Cl NMR data demonstrate 
a pseudo spin-gap behavior below $T^{*}=135 K$, followed by successive phase transitions 
at $T_{c1}=94 K$ and $T_{c2}=66 K$ into the singlet ground 
state with an unusually large energy gap $E_{g}=430 K$.  
The broad distribution of the local electric field gradient in the 
intermediate phase between $T_{c1}$ and $T_{c2}$ suggests 
unconventional spin and lattice (probably orbital) states with an emerging spin-gap.
\end{abstract}
\pacs{76.60.-k, 75.10.Jm}
}
The provocative proposal that the mechanism of high temperature superconductivity in 
layered cuprates may be related to the exotic properties of low dimensional 
quantum spin systems, such as the RVB (Resonanting Valence Bond) state \cite{Anderson}, 
has been a major driving force behind the rapid advance of {\it quantum 
magnetism}.  Naturally, the initial emphasis was 
placed on understanding
materials involving Cu$^{2+}$ ions with the 3d$^{9}$ configuration
($S=\frac{1}{2}$) \cite{Hase,Azuma}.  More recently, 
$S=\frac{1}{2}$ quantum magnets involving V$^{4+}$ \cite{Ueda1,Ueda2,Ueda3}
and Ti$^{3+}$ \cite{Tokura,Keimer,Ueda4,Ueda5,Beynon,Seidel} ions with 3d$^{1}$ 
configurations have been attracting strong attention.  Potential interests 
include the realization of doped metallic states 
in 3d$^{1}$ {\it Mott insulators}, and ultimately, superconductivity\cite{Ueda3}.  
In addition, the near degeneracy of the $t_{2g}$-orbitals 
often gives rise to the orbital and/or charge ordering \cite{Nagaosa}.  In the orbital 
ordered state, the configuration of the occupied 3d$_{xy,yz,zx}$-orbitals exhibits a long range order.  
On the other hand, the valence of ions differs site by site in charge 
ordered states.  The additional orbital and charge degrees of 
freedom make the underlying physics of 3d$^{1}$ quantum spin systems more intriguing, yet more complicated.  
A fascinating example is a mixed valence system NaV$_{2}$O$_{5}$ 
\cite{Ueda2,Ohama1} (the 
average valence of V-ions is +4.5).  NaV$_{2}$O$_{5}$ undergoes 
orbital and charge ordering at  $T_{c}=34K$, where a large energy gap 
$E_{g}=98K$ opens \cite{Ohama1,Fagot,Ohama2,Yoshihama}.  The ratio 
$2E_{g}/k_{B}T_{c}=6$ is much larger than the BCS value 3.5 
expected for conventional spin-Peierls transitions with lattice-dimerization.  
Intensive theoretical and experimental efforts have been underway to 
understand the exotic phase transition.

An equally exciting new avenue to investigate $S=\frac{1}{2}$ quantum spin 
systems with 3d$^{1}$ electrons is the titanates 
involving Ti$^{3+}$ ions.  Starting from 
the metal-insulator transition
in Sr$_{1-x}$La$_{x}$TiO$_{3}$\cite{Tokura}, growing efforts are 
under way in search for a new 
form of quantum magnetism in titanates: strong orbital fluctuations in LaTiO$_{3}$\cite{Keimer};
 a quasi 
1D $S=\frac{1}{2}$ 
chain system (Na,Li)TiSi$_{2}$O$_{6}$ \cite{Ueda4}; 3D Pyrochlore 
system MgTi$_{2}$O$_{4}$\cite{Ueda5}; and
a quasi-2D layered system 
TiOCl\cite{Beynon,Seidel}.   

Among these titanates, the TiOCl system has particularly unique characteristics.  
First, Ti$^{3+}$O$^{2-}$ form bi-layers separated by 
Cl$^{-}$ bi-layers.  The quasi-2D layered structure may be considered 
a close analogue to those realized in high $T_{c}$ 
cuprates.  In fact, Beynon and Wilson \cite{Beynon}
reported the very little temperature dependence in uniform 
susceptibility $\chi(T)$, 
and discussed the possible realization of a RVB ground 
state in the Mott-insulator.  
They also noted that $\chi(T)$ is very sensitive to 
impurities, and demonstrated that a Sc$^{3+}$ ion ($S=0$) substituted into a Ti$^{3+}$ 
site ($S=\frac{1}{2}$) gives rise to a localized spin 
$S=\frac{1}{2}$, in 
analogy with creation of a free spin by Zn$^{2+}$ ($S=0$) substituted into Cu$^{2+}$ ($S=\frac{1}{2}$) 
sites in cuprates.  Second, in a very recent report, 
Seidel et al. observed a sharp, nearly isotropic drop of $\chi(T)$ below 
$\sim100K$ in defect-free samples, signaling the emergence of an energy 
gap\cite{Seidel}.  The drop in $\chi(T)$ is even more pronounced in high quality single 
crystals.  Since $\chi(T)$ above $\sim200K$ can be 
fitted with a $S=\frac{1}{2}$ Heisenberg chain model with the nearest neighbor 
exchange interaction $J=660K$, 
Seidel et al. proposed, based on LDA+U calculations, that 
the effective dimension of the TiO layers are reduced from 2D to 1D by
an orbital order at Ti$^{3+}$ sites along the a- or b-axis.  The gapped behavior of $\chi(T)$ 
was attributed to a spin-Peierls transition \cite{Seidel}.  Third, 
but related to the 
second point, the near degeneracy of the energy levels of 
different 3d$^{1}$ orbital configurations {\it without mixed-valence 
nature} might make TiOCl an ideal model system to investigate 
spin-Peierls-{\it like} 
transitions with additional orbital degrees of freedom but probably without 
charge order.

In this Letter, we report the first NMR investigation of TiOCl.  While the 
spin-Peierls transition in CuGeO$_{3}$ ({\it without} orbital and 
charge degrees of freedom) and 
the exotic order in NaV$_{2}$O$_{5}$ ({\it with} orbital {\it and} charge degrees of freedom) have seen 
many detailed microscopic investigations, to the best of our 
knowledge this is the first successful microscopic experiments reported for TiOCl.  
We demonstrate that TiOCl reveals a unique 
spin-gap behavior accompanied by lattice instabilities, and undergoes successive 
phase transitions at $T_{c1}=94\pm2K$ and $T_{c2}=66\pm2 K$.
Unlike CuGeO$_{3}$ and NaV$_{2}$O$_{5}$, the fluctuation effects in 
TiOCl are so strong above $T_{c1}$ that
a pseudo spin-gap 
manifests itself as high as $T^{*}=135\pm10K$.  Moreover, our 
observation of a broad continuum in the NMR 
lineshape data indicates that the intermediate phase between $T_{c1}$ 
and $T_{c2}$ is {\it not} a simple, dimerized state in 1D.  Below 
$T_{c2}$, TiOCl undergoes a first order phase transition to 
a fully gapped state (rather than to the gapless RVB state) with an 
extraordinarily 
large energy gap $E_{g}=430\pm60K$.  The
unusually large energy gap ($2E_{g}/k_{B}T_{c1,c2}=10\sim 15$) as well 
as the presence of the intermediate state between $T_{c1}$ and 
$T_{c2}$ suggest that the observed 
phase transitions are not conventional spin-Peierls transitions, and 
point towards the significance of the roles played by the additional orbital 
degrees of freedom over the entire temperature range.  

Our TiOCl single crystals were synthesized by standard vapor-transport 
techniques from TiO$_{2}$ and 
TiCl$_{3}$\cite{Seidel}.   A large number of very thin, flaky single crystals
with typical dimensions of 2mm$\times$2mm were assembled on a Macor 
sample holder.  NMR measurements were conducted by applying an 
external magnetic field in parallel with the aligned crystal c-axis.  
Random alignment within the TiO-plane prevented us from conducting 
measurements 
along the a- and b-directions.  We emphasize that, unlike $^{63,65}$Cu and $^{51}$V 
NMR, the sensitivity of $^{47,49}$Ti NMR is notoriously low 
\cite{sensitivity}.  Very short transverse relaxation times $T_{2}$ at Ti 
sites and the small volume of the available crystals made the NMR 
measurements even more difficult.  We needed to average spin-echo 
signals up to $\sim10^{6}$ scans to obtain a reasonable signal to noise 
ratio.  In general, NMR signal intensities 
increase as $1/T$ in proportion to the Boltzman 
factor.  However, the exponentially growing spin-lattice relaxation time $T_{1}$ below 
$T_{c1}$ slows down the NMR pulse-sequence, hence 
the measurements at lower temperatures were equally difficult and time 
consuming.  Despite our intensive efforts, we have been able 
to find $^{47,49}$Ti NMR signals only for the central transition 
from the $I_{z}=+\frac{1}{2}$ to 
$-\frac{1}{2}$ state.  The detection of $^{35}$Cl NMR signal was 
somewhat easier, and we did manage to find the $I_{z}=\pm\frac{3}{2}$ to 
$\pm\frac{1}{2}$ transitions above $T_{c1}$ \cite{nuq}.  All the nuclear 
spin-lattice relaxation data were deduced by fitting the nuclear spin 
recovery after an inversion $\pi$-pulse to the standard rate equations.  
We confirmed that $^{47}$Ti (nuclear spin $I=5/2$) and $^{49}$Ti (nuclear 
spin $I=7/2$) NMR gives identical $1/T_{1}$.  Given that the magnetic recovery 
process of nuclear magnetization $M(t)$ is dominated by different 
terms for $^{47}$Ti ($M(t)\sim exp[-15t/T_{1}]$) and $^{49}$Ti ($M(t)\sim exp[-28t/T_{1}]$), 
we conclude that $1/T_{1}$ is dominated entirely by magnetic fluctuations at 
all temperatures.  The difference in the recovery characteristic of 
$M(t)$ also helped us identify the pairs of $^{47}$Ti  and 
$^{49}$Ti lines below $T_{c2}$, where the NMR lines split into 
doublets.    

The experimental information on the spin degrees of freedom is 
summarized in Fig. 1.  Quite generally, the $^{47,49}$Ti
nuclear spin-lattice relaxation rate, caused by low frequency 
magnetic fluctuations may be expressed as\cite{Moriya},
\begin{equation}
          \frac{1}{T_{1}} = T\frac{ k_{B}\gamma_{n}^{2}}{\mu_{B}^{2} \hbar} 
	\sum_{{\bf q}} | F({\bf q}) |^{2} \frac{\chi''({\bf 
	q},\omega_{n})}{\omega_{n}},
\label{T1}
\end{equation}
where $F({\bf q})$ is the form factor of the electron-nucleus hyperfine 
couplings, and $\chi''({\bf q},\omega_{n})$ is the 
imaginary part of the dynamical susceptibility at the 
observed NMR frequency $\omega_{n}$.  $1/T_{1}$ appears to asymptote to a 
constant value near $300K$, $1/T_{1}\sim 1400$ sec$^{-1}$.  This implies 
that $\chi''({\bf q},\omega_{n})$ asymptotes to a Curie law at 
higher temperatures.  This is consistent with the behavior expected for 
typical $S=\frac{1}{2}$ 1D 
Heisenberg chains such as Sr$_{2}$CuO$_{3}$ 
\cite{Takigawa}. 
However, we caution that $1/T_{1}\sim const.$ may also be expected
for $S=\frac{1}{2}$ 2D Heisenberg model\cite{Imai93}.  Certainly $\chi(T)$ of 
TiOCl also fits nicely 
to the 1D Heisenberg chain model between $200K$ and $800K$ \cite{Seidel}, but we 
recall that NaV$_{2}$O$_{5}$ turns out to be a 2D spin-charge-orbital 
hybrid system \cite{Gaulin} despite the similarly nice fit of $\chi(T)$ 
to the 1D model.  

In general, the growth of short-range order enhances low frequency 
spin fluctuations (hence $1/T_{1}T$) with 
decreasing temperature in the absence of a gap in the spin excitation 
spectrum.  The most striking feature in Fig.1 is that 
$1/T_{1}T$ begins to {\it decrease} below $T^{*}\sim135K$.  This implies 
that low frequency spin fluctuations are suppressed below 
$T^{*}$, by almost two orders of magnitude 
 between $135K$ and $65K$.  Interestingly, the observed behavior of $1/T_{1}T$ 
is qualitatively similar to the pseudo-gap phase in underdoped 
high $T_{c}$ cuprates.  Needless to say, it does not necessarily imply 
that the underlying mechanism is identical.  

We can also probe the lattice degrees of freedom by observing the EFG (Electric Field Gradient) 
reflected on NMR lineshapes.  
In Fig.2, we present the $^{47,49}$Ti and $^{35}$Cl NMR lineshapes for 
the central transition.  We observed a single NMR line for 
both Ti and Cl down to $T_{c1}=94\pm2 K$.  For $^{35}$Cl, we also 
managed to detect $I_{z}=\pm\frac{3}{2}$ to $I_{z}=\pm\frac{1}{2}$ 
satellite transitions, but again there was only one kind of signal \cite{nuq}.   These results indicate that there is only 
one kind of Ti and Cl site in TiOCl within our experimental resolution, and rule out 
any potential orbital order configurations above $T_{c1}$ that would lead 
to more than one inequivalent sites.  Below $T_{c1}$, both $^{47,49}$Ti and 
$^{35}$Cl NMR central lines begin to broaden, signaling a second order phase 
transition.  We confirmed that the magnitude of the $^{35}$Cl NMR line splitting is 
inversely proportional to the external magnetic field.  Therefore the line splitting is caused by the second-order 
nuclear quadrupole interaction with the EFG.  We also found that the drop of $1/T_{1}T$ is 
accelerated below $T_{c1}$.
In addition, close inspection of the $\chi(T)$ 
data reported in \cite{Seidel} reveals a kink at $T_{c1}$, followed by a rapid decrease 
with temperature with positive curvatures.  These results suggest that 
a clear gap structure emerges in the spin excitation spectrum at $T_{c1}$, accompanied 
by a {\it static} distortion of the lattice.  The NMR lineshape exhibits a broad continuum below 
$T_{c1}$ down to $T_{c2}$ as summarized in Fig.3(b).
The broad continuous distribution of the local EFG environment
implies the presence of numerous inequivalent Ti and Cl sites in the TiOCl 
lattice.  The most plausible scenario is that the emergence of the 
spin-gap at $T_{c1}$ is accompanied by an orbital-order (possibly 
incommensurate).  That is, the phase transition 
at $T_{c1}$ is {\it not} a simple spin-Peierls transition due to lattice 
dimerization, even though the well-developed short-rage spin order below 
$400 K$ and the gapped behavior of $\chi(T)$ below $\sim100K$ \cite{Seidel} suggest the contrary.

We found additional evidence for the involvement of the lattice in the phase 
transition at $T_{c1}$ by $^{35}$Cl nuclear relaxation data $^{35}1/T_{1}$ as shown 
in Fig.3(a).  $^{35}1/T_{1}$ is two 
orders of magnitude slower than $1/T_{1}$ at $^{47,49}$Ti-sites.  
This implies that the magnetic hyperfine coupling of $^{35}$Cl nuclear 
spins with Ti 3d$^{1}$ electron spins is at least by an order of 
magnitude smaller.
The qualitatively different temperature dependence between  
$^{35}1/T_{1}$ [see Fig.3(a)] and $1/T_{1}$ [see Fig.1(a)]
suggests that slowing of EFG fluctuations below $T^{*}$
toward the phase transition at $T_{c1}$ causes the cusp of $^{35}1/T_{1}$ 
at $T_{c1}$\cite{quadrupole}.  
Even more straightforward evidence is that the $^{35}$Cl  spin-echo decay (transverse 
$T_{2}$ relaxation) in Fig.3(c) and (d) shows typical motional narrowing effects below 
$T^{*}$ towards $T_{c1}$ due to lattice softening.  
The slow lattice dynamics at or below the $^{35}$Cl NMR frequency 
(37.6 MHz) transforms the spin echo-decay to Lorentzian-like as $T_{c1}$ 
is approached.  Taken together with the precursive suppression observed for $1/T_{1}T$ 
below $T^{*}$ in Fig.1(a), we conclude that the drastic softening of the lattice below 
$T^{*}$ drives the observed pseudo spin-gap behavior
prior to the actual second-order phase transition at $T_{c1}=94K$. 
The high onset temperature $\sim200K$ of the gradual decrease of 
$1/T_{1}$ suggests that phonon softening at higher frequency scales begins at much higher 
temperature. 

At $T_{c2}$, the continuum of the $^{35}$Cl and $^{47,49}$Ti NMR 
lineshape suddenly collapses into doublets, whose line positions 
correspond to the two extrema observed above $T_{c2}$.  This strongly 
suggests that the unit cell of TiOCl has two and only two inequivalent 
Ti and Cl sites below $T_{c2}$ \cite{tensor}.  The low-frequency Ti 3d$^{1}$ spin 
fluctuations exhibit an activation behavior below 
$T_{c2}$, as shown in Fig.1(b).  From the fit to an exponential form, $1/T_{1}T\sim 
exp(-E_{g}/k_{B}T)$, we deduce the energy gap $E_{g}=430\pm60K$.  
These results are consistent with the formation of a singlet ground 
state at $T_{c2}$ by lattice dimerization.  Our results are also consistent with the recent 
observation of doubling of the unit cell
along the b-axis at $66\pm1K$ by Y.S. Lee et al. based 
on x-ray scattering measurements\cite{x-ray}.  On the other hand, 
there are many experimental signatures which are at odds with conventional 
second order spin-Peierls 
transitions.  $1/T_{1}$ [Fig.1], the NMR lineshapes [Fig.2], 
and $^{35}1/T_{1}$ [Fig.3(a)] change discontinuously at $T_{c2}$.  
$^{35}$Cl spin-echo decay curve also changes suddenly to a slow 
Gaussian type at $T_{c2}$ [Fig.3(d)].  This indicates that the soft 
vibration of the lattice disappears suddenly below $T_{c2}$.  
All these NMR results as well as the discontinuous change in $\chi(T)$ 
\cite{Seidel} and the x-ray data \cite{x-ray} at $T_{c2}$ indicate that the phase 
transition at $T_{c2}$ is first order.  Furthermore, the observed energy gap is 
unusually large, $2E_{g}/k_{B}T_{c1,c2}=10\sim 15$.  
In the case of conventional, mean-field spin-Peierls gap, one 
would expect the ratio to be close to the BCS value, $2E_{g}/k_{B}T_{c}=3.5$, 
as observed for CuGeO$_{3}$ ($2E_{g}/k_{B}T_{c}=3.3$\cite{Kikuchi,Regneau}).  
The observed ratio in TiOCl is even larger than that for NaV$_{2}$O$_{5}$, 
$2E_{g}/k_{B}T_{c}\sim 6$ \cite{Ohama1,Yoshihama}. 
It is important to realize that $E_{g}$ is comparable to the apparent 
exchange energy $J=660 K$ estimated by the
1D fit of $\chi(T)$ \cite{Seidel}.  Such a large magnitude of the energy gap $E_{g}$ 
strongly suggests that the spin excitations from the singlet ground state 
is dressed by other electronic degrees of freedom, most likely of the 
orbital origin.

To summarize, TiOCl is a rather unique layered
$S=\frac{1}{2}$ quantum-material with a pre-existing pseudo spin-gap above 
$T_{c1}$, the unconventional intermediate spin and lattice 
(probably orbital) states between $T_{c1}$ 
and $T_{c2}$, and a first order phase transition into a singlet ground state 
with an unusually large energy gap.  
At first sight, the strong fluctuation effects evidenced in the pseudo spin-gap 
behavior suggest a genuinely 1D nature of the TiOCl along the a- or b-axis, 
achieved by an orbital 
order well above $T_{c1}$.  We recall that the pre-existing orbital order above $T_{c1}$ may 
reduce the effective dimensions of the TiO layer from 2D to 1D, as 
suggested by Seidel et al. based on LDA+U calculations\cite{Seidel}.  
On the other hand, the additional orbital 
degrees of freedom along the two orthogonal directions within the 
2D TiO-layers may cause strong orbital fluctuations, which could effectively
 suppress the tendency towards a spin-Peierls transition at finite 
temperature.  In such a scenario, the manifestation of the pseudo spin-gap 
above $T_{c1}$ may be the consequence of the suppression of the spin-Peierls 
transition, and the successive phase transitions at $T_{c1}$ and 
$T_{c2}$ may arise from a competition between different 
orbital orders.  In addition, if the second nearest-neighbor exchange interaction 
$J_{a}'$ in the zig-zag chain structure along the a-axis (see Fig.1(c) 
in \cite{Seidel}) exceeds 
$0.241J_{a}$ (where $J_{a}$ being the nearest-neighbor exchange interaction along 
the a-axis), a spontaneous spin dimerization along the a-axis may be favoured 
\cite{second} over the unit-cell doubling along the b-axis observed below 
$T_{c2}$, introducing frustration.  
In any case, the unprecedented behavior of TiOCl points 
towards the crucial roles played by the orbital degrees 
of freedom and their fluctuations below $\sim200K$.    
Recalling that TiOCl is not a mixed-valence system such as 
NaV$_{2}$O$_{5}$,  
TiOCl may be an ideal model spin-Peierls system with additional 
orbital degrees of freedom, and we call for futher microscopic studies.

We thank P.A. Lee, Y.S. Lee, G. Sawatzky, V. Kiryukhin, and B.D. Gaulin for helpful discussions, 
and I. Affleck for calling our attention to Ref. \cite{second}.
The work at M.I.T. was supported by NSF DMR 98-08941.


%
%

\begin{figure}
\caption{(a) $1/T_{1}$ ($\circ$) and 
$1/T_{1}T$ ($\bullet$) at $^{47,49}$Ti sites.  
$1/T_{1}$ was measured at the center peak for $T \lesssim T_{c1}$.
Solid and dashed curves are guides for 
eyes.  (b) The same $1/T_{1}T$ data plotted in a 
semi-log scale.  Solid line is the best exponential fit.  
$1/T_{1}T$ was identical for the doublets below $T_{c2}$.
}
\label{Fig1.eps}
\end{figure}

\begin{figure}
\caption{NMR lineshapes of the $I_{z}=+\frac{1}{2}$ to 
$-\frac{1}{2}$ central transition for $^{35}$Cl 
(left panels) and $^{47,49}$Ti (right panels).  The FFT 
envelope was obtained at 9 Tesla by repeating spin-echo measurements at several 
frequencies.}
\label{Fig2.eps}
\end{figure}

\begin{figure}
\caption{(a) $^{35}1/T_{1}$.
(b) $^{35}$Cl NMR frequencies observed at 9 Tesla.  Dashed lines 
represent continuum.  Solid error bars represent the frequency range 
corresponding to the half-intensity.  (c) $^{35}$Cl 
NMR spin-echo decay observed at the main peak, as a function of the pulse separation time $2\tau$ 
between 90-180 degree R.F. pulses above $T_{c1}$, and (d) below 
$T_{c1}$.  Solid curves are guides for eyes. }
\label{Fig3.eps}
\end{figure}

%
%


\begin{references}
\bibitem{sensitivity}
The natural abundance of $^{47,49}$Ti is low (7.3\% and 5.5\%, 
respectively).  The nuclear gyromagnetic ratio is also rather small 
($^{47,49}\gamma_{n}/2\pi=2.4000$ and $2.4005$ MHz/Tesla, 
respectively).
\bibitem{nuq}The $^{35}$Cl quadrupole tensor along 
the c-axis was determined to be $^{35}\nu_{Q}^{c}=1.641\pm 0.005 MHz$ 
at 130K.
\bibitem{quadrupole}Strictly speaking, fitting the $^{35}$Cl nuclear spin 
recovery $M(t)$ to a purely magnetic form may not be valid near $T_{c1}$.  
However, it is sufficient for the present purpose.
\bibitem{tensor}Our preliminary attempts to deduce more detailed information 
on the lattice distortion at $T_{c1}$ and $T_{c2}$ have been 
unsuccessful, as NMR signals are too weak to detect satellite 
transitions below $T_{c1}$.  We defer further efforts until much larger single crystals become 
available.
\end{references}
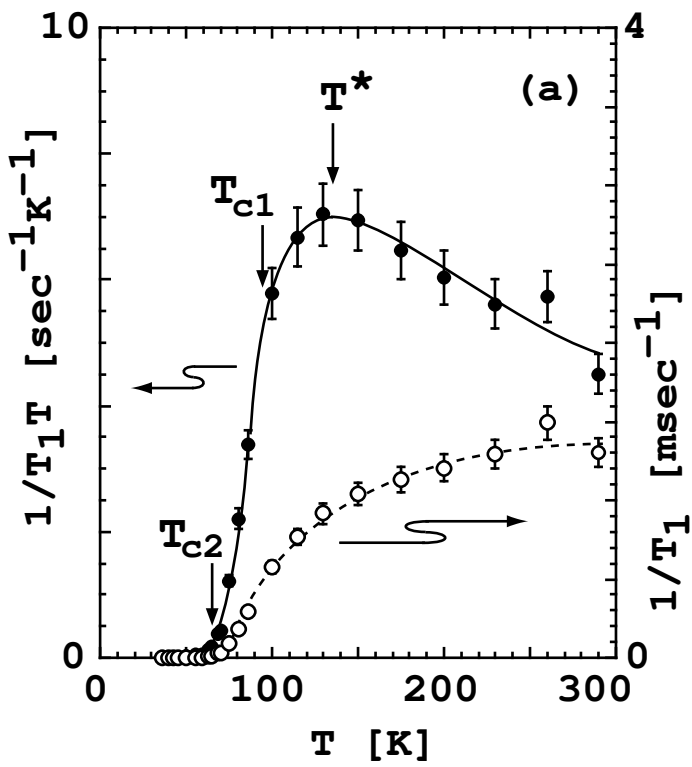
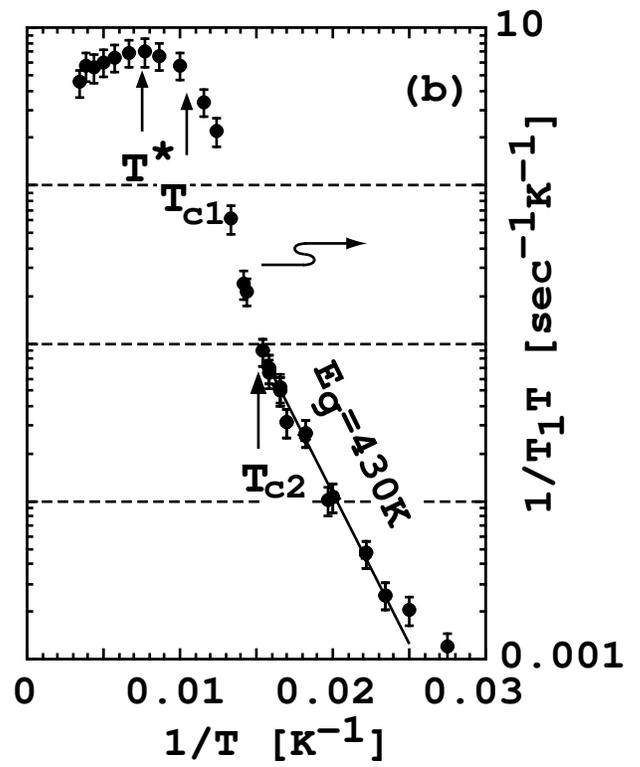
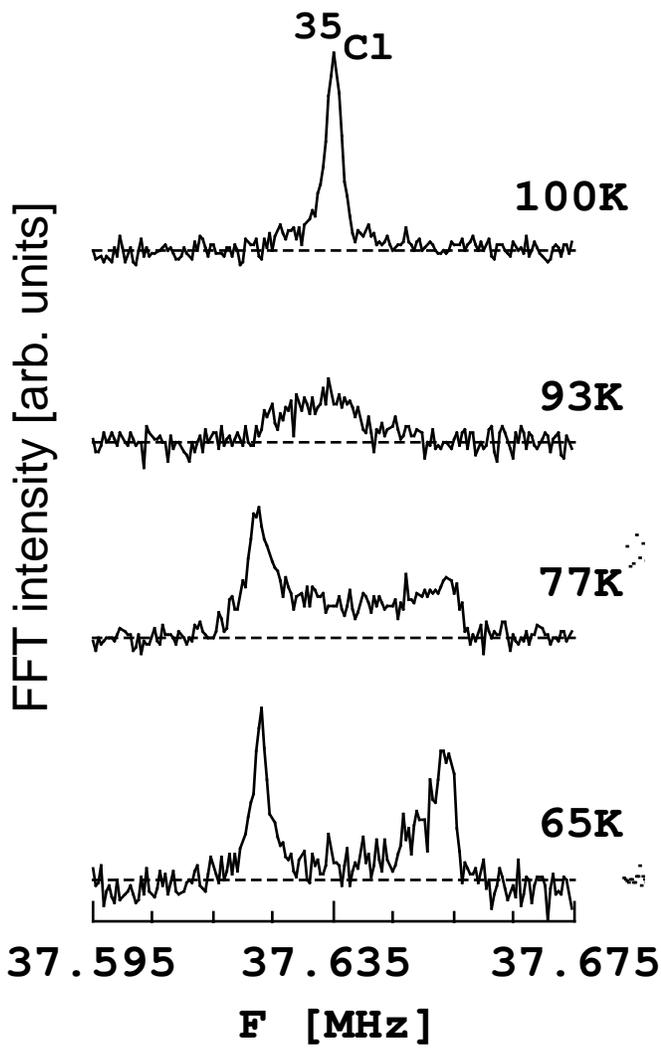
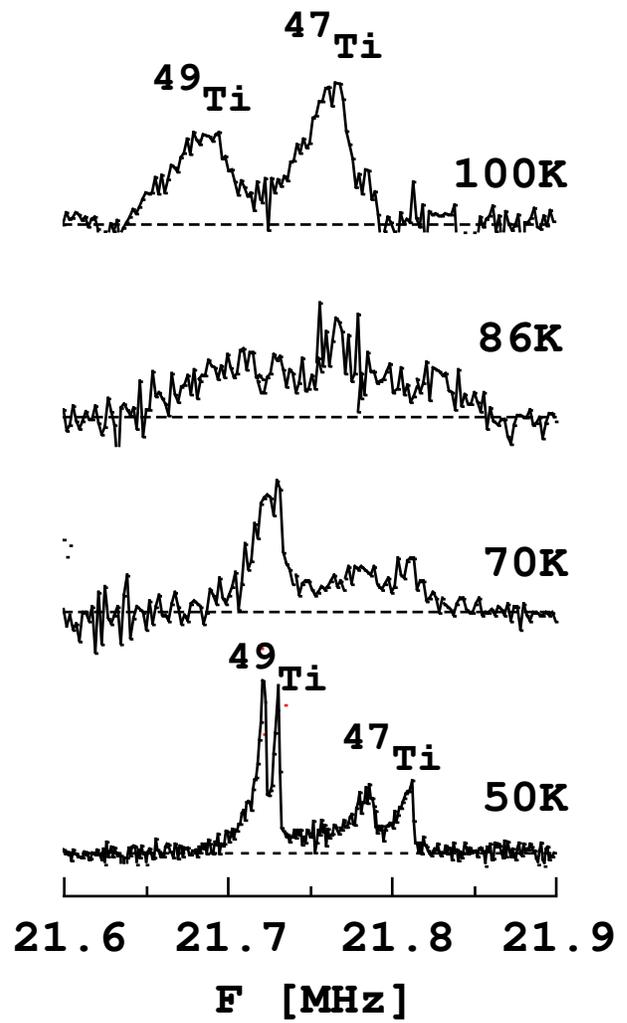
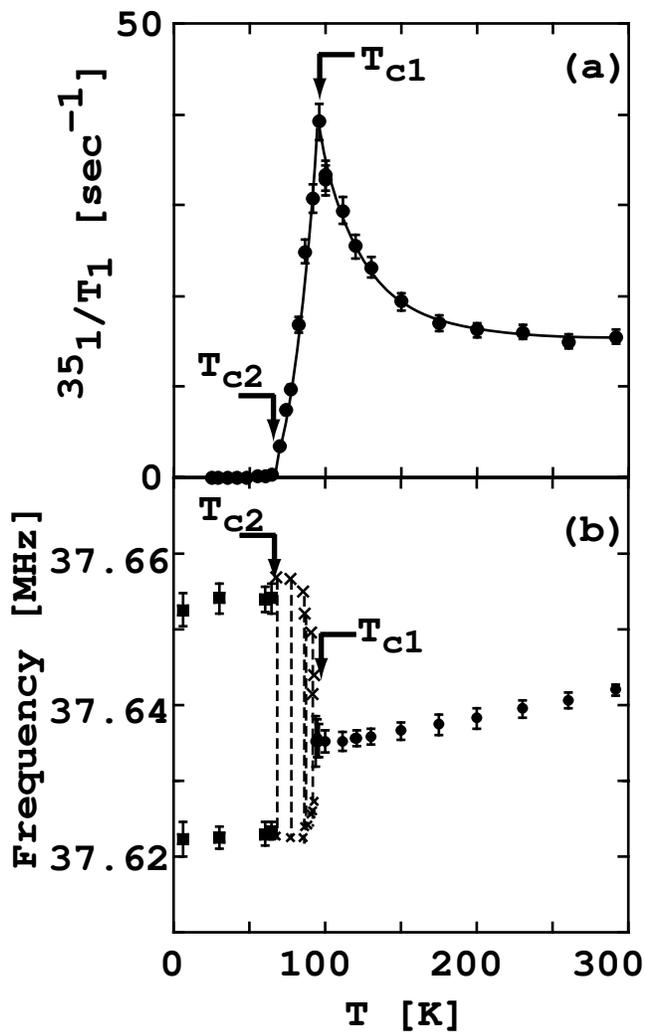
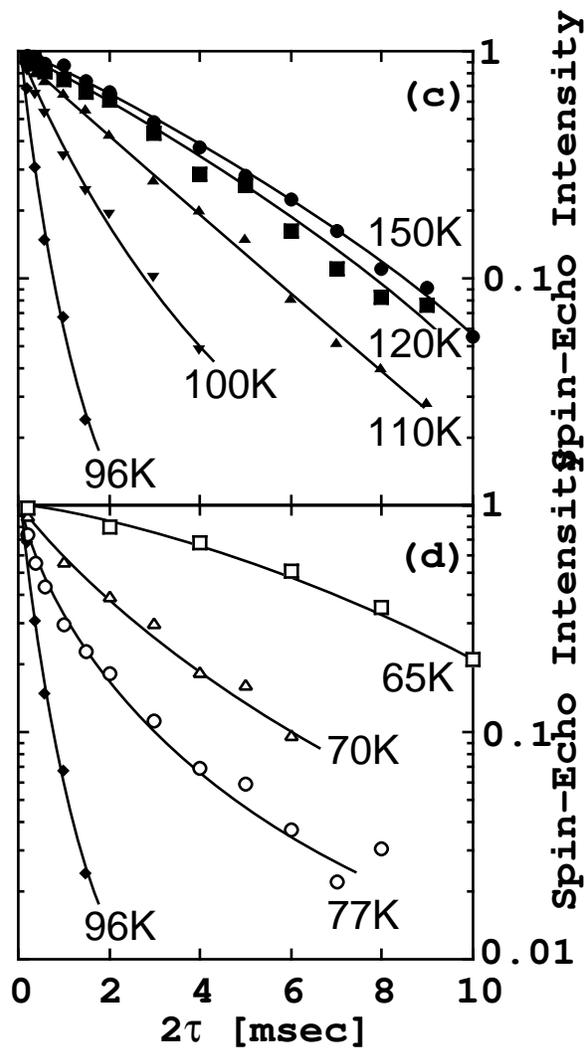
\end{document}